\documentclass[reprint,
superscriptaddress,
 amsmath,amssymb,
 aps,pra,
showkeys,
]{revtex4-2}

\usepackage[utf8]{inputenc}
\usepackage{longtable}
\usepackage{morefloats}
\usepackage{tabularx}
\usepackage{multirow}
\usepackage{graphicx}
\usepackage{color}
\usepackage{epsfig,amsbsy}
\usepackage{amsmath,amsfonts,amsthm,amssymb}
\usepackage{appendix}
\usepackage{textcomp}
\usepackage{makeidx}
\usepackage{epstopdf}
\usepackage[normalem]{ulem}
\usepackage[bookmarksnumbered,pdfpagelabels=true,plainpages=false,colorlinks=true,linkcolor=blue,citecolor=red,urlcolor=blue]{hyperref}

\usepackage{soul}

\usepackage{xcolor}

\NewDocumentCommand{\fhighlight}{O{blue!40} m m}
{\draw[rounded corners,ultra thin, fill=red,draw=black, opacity=0.1] (m-1-1.north west)rectangle (m-2-2.south east);
\draw[rounded corners,ultra thin, fill=red,draw=black, opacity=0.1] (m-3-3.north west)rectangle (m-4-4.south east);
\draw[rounded corners,ultra thin, fill=blue,draw=black, opacity=0.03] (m-1-3.north west)rectangle (m-2-4.south east);
\draw[rounded corners,ultra thin, fill=blue,draw=black, opacity=0.03] (m-3-1.north west)rectangle (m-4-2.south east);}

\begin{document}

\title{Chiral Tubular Magnonic Crystals with Periodic Dzyaloshinskii--Moriya Interaction}

\author{P. Contreras-Gallardo}
\affiliation{Departamento de F\'isica, Universidad T\'ecnica Federico Santa Mar\'ia, Avenida Espa\~na 1680, Valpara\'iso, Chile}

\author{J. Flores-Far\'{i}as}
\affiliation{Departamento de F\'{i}sica, Facultad de Ciencias F\'{i}sicas y Matem\'{a}ticas, Universidad de Chile, Santiago, Chile}

\author{B. Mimica-Figari}
\affiliation{Department of Electrical and Information Engineering, Politecnico di Bari, Bari 70126, Italy}

\author{V. Puliafito}
\affiliation{Department of Electrical and Information Engineering, Politecnico di Bari, Bari 70126, Italy}

\author{P. Landeros}
\affiliation{Departamento de F\'isica, Universidad T\'ecnica Federico Santa Mar\'ia, Avenida Espa\~na 1680, Valpara\'iso, Chile}

\author{R. A. Gallardo}
\email{rodolfo.gallardo@usm.cl}
\affiliation{Departamento de F\'isica, Universidad T\'ecnica Federico Santa Mar\'ia, Avenida Espa\~na 1680, Valpara\'iso, Chile}

\date{\today }

\begin{abstract}
We introduce a chiral tubular magnonic crystal formed by a ferromagnetic nanotube with a periodically modulated interfacial Dzyaloshinskii--Moriya interaction (DMI). Using a plane-wave formalism adapted to the cylindrical geometry, we calculate the spin-wave band structure including exchange, nonlocal dipolar coupling, and interfacial DMI. We show that the tubular geometry qualitatively modifies the role of periodic DMI compared with planar chiral magnonic crystals. Besides producing the usual nonreciprocal frequency shift, curvature converts part of the interfacial DMI into an effective anisotropy-like internal field. This creates a sign-dependent frequency landscape along the tube, so that DMI-covered regions can act either as potential wells or barriers for the lowest-frequency modes. As a result, flat and weakly dispersive bands appear in both Damon--Eshbach-like and backward-volume-like configurations, with the latter having no direct counterpart in planar systems. Our results identify periodically engineered DMI in nanotubes as a route for controlling nonreciprocal spin waves, localized modes, and flat magnonic bands in curved architectures.
\end{abstract}

\maketitle

\section{Introduction}
\label{sec:introduction}

Spin waves, or magnons, are collective excitations of the magnetic order that can propagate at GHz frequencies with wavelengths far below those of electromagnetic waves in the same spectral range \cite{Neusser09,Grundler16,Tacchi17rev,Petti22}. This makes them attractive information carriers for nanoscale devices, where signal transmission can be achieved without charge transport, thereby reducing Joule losses \cite{Gurevich96,Rezende20,Chumak22}. A central goal in magnonics is to control the dispersion, propagation direction, group velocity, and spatial localization of these excitations by engineering the magnetic energy landscape \cite{Davies15,Chen20,Barman21,Flebus24}.

A versatile route toward spin-wave control is provided by magnonic crystals, i.e., artificial magnetic media in which geometric or material parameters are periodically modulated \cite{Krawczyk08,Wang10,Mruczkiewicz14,Klos16,Langer17,Flores25b,Flores26c}. In such metamaterials, spin waves form band structures with allowed and forbidden frequency ranges, analogous to photonic and phononic crystals \cite{Kruglyak10,Lenk11,Demokritov12}. The resulting band gaps, defect modes, localized states, and flat bands provide mechanisms for filtering, guiding, trapping, and slowing down spin-wave excitations \cite{Gubbiotti05,Zhang12,Pan13,Di14a,Gallardo18a,Gallardo18bb,Grassi20,Cheng23,Li26,Chen26}. Magnonic crystals can be realized through periodic variations of the saturation magnetization, exchange stiffness, anisotropy, thickness, or interlayer coupling, thus offering a broad platform for tailoring the spin-wave spectrum \cite{Krawczyk13,Chumak17,Alvarado19}.
They can also be realized in simple magnetic films hosting a periodic magnetic texture \cite{Kugler15,Garst17,Yu21,Timofeev22,Ogawa2021}. 

Beyond conventional periodic modulations of material or geometric parameters, chiral magnetic interactions offer an additional route to engineering magnonic band structures. A prominent example is the Dzyaloshinskii--Moriya interaction (DMI) \cite{Dzyaloshinskii58,Moriya60,Fert80,Crepieux98}, an antisymmetric exchange interaction arising from spin-orbit coupling in systems with broken inversion symmetry. The DMI favors noncollinear spin arrangements \cite{Roszler06,Nagaosa13,Schwarze15,Finocchio16,Jena20,Gobel21,Kisielewski23,Cepeda26} and induces nonreciprocal spin-wave propagation \cite{Udvardi09,Cortes13,Cho15,Nembach15,Seki16,Iguchi15,Tacchi17,Gallardo19BCh,Kuepferling23,Camley23}.
In ultrathin ferromagnetic (FM) films covered with a heavy-metal (HM) layer, the strength of the interfacial DMI can be locally controlled by varying the thickness of the HM layer \cite{Tacchi17}. 
This idea led to the proposal of chiral magnonic crystals with a periodic DMI, in which the antisymmetric exchange itself serves as the spatially modulated magnonic potential \cite{Gallardo19c}. In such planar systems, the periodic DMI produces unconventional features such as indirect band gaps, low-frequency flat bands, and spin-wave modes localized where the DMI is active \cite{Gallardo19c,Flores22,Flores24}. These predictions have been confirmed experimentally in one-dimensional chiral magnonic crystals, where flat bands and asymmetric spin-wave amplitudes were observed by Brillouin light scattering \cite{Tacchi23,Wei25a}. Subsequent theoretical and experimental works extended this concept to two-dimensional lattices, chiral superlattices, quasiperiodic arrangements, and graded-DMI systems, demonstrating the robustness and tunability of DMI-induced localization and nonreciprocity \cite{Flores25,Wei25b,Flores26,Flores26b}.

At the same time, curved and three-dimensional magnetic systems have emerged as a distinct route for controlling magnetization dynamics \cite{Sheka22,Sheka22b,Gubbiotti25}. In curvilinear shells, the geometry itself modifies the magnetic interactions and may generate effective chiral terms, curvature-induced anisotropies, and direction-dependent dipolar couplings \cite{Landeros10,Otalora12,Yan12,Gaididei14,Sheka15,Stano18,Makarov21,Sheka21,Landeros22}. Magnetic nanotubes are particularly relevant in this context because their closed cylindrical geometry supports axial, vortex, and curling magnetization states, each producing distinct spin-wave symmetries and selection rules \cite{Otalora16,Sheka20,Salazar21,Gallardo22a,Korber21b,Korber22,Brevis24,Brevis25,Mimica25,Mimica26}. For vortex or curling configurations, the interplay between curvature, dipolar fields, exchange interaction, and equilibrium chirality can lead to nonreciprocal axial propagation, magnetochiral effects, and mode asymmetries that have no direct counterpart in planar films \cite{Brevis25,Mimica25,Mimica26}. Moreover, when a nanotube is interfaced with a heavy metal, interfacial DMI provides an additional chiral mechanism that can either reinforce or compete with the curvature-induced nonreciprocity, depending on the magnetic state and the DMI sign \cite{Mimica26}.

Despite these advances, periodic interfacial DMI has so far been explored only in planar magnonic crystals and superlattices. The possibility of combining a spatially modulated DMI with the intrinsic curvature of a tubular shell remains unexplored. This raises several fundamental questions. How does a periodic chiral interaction modify the spin-wave band structure of a nanotube? Can DMI-induced flat bands and band gaps persist in a cylindrical geometry? How do curvature and tubular boundary conditions reshape the localization
and nonreciprocity mechanisms known from planar chiral magnonic crystals?

In this work, we address these questions by introducing a chiral curved magnonic crystal, consisting of a magnetic nanotube with a periodically modulated interfacial Dzyaloshinskii--Moriya interaction. For instance, the modulation may be realized by covering the nanotube with a periodic array of heavy-metal rings that induce DMI beneath selected regions of the FM cylindrical shell. We develop a plane-wave formalism adapted to the tubular geometry, in which the dynamic magnetization and the spatially dependent DMI are expanded in reciprocal-space harmonics consistent with the axial periodicity. This approach allows us to calculate the spin-wave band structure while retaining the combined effects of exchange, dipolar coupling, curvature, and interfacial DMI.
We show that the tubular geometry qualitatively modifies the role of periodic DMI compared with planar chiral magnonic crystals. In addition to producing the usual nonreciprocal frequency shift, the interfacial DMI acquires a curvature-dependent contribution that enters the spin-wave dispersion as an effective anisotropy field. As a consequence, the DMI-active and DMI-free regions form a sign-dependent internal-field landscape, which determines whether the lowest-frequency modes localize inside or outside the heavy-metal-covered regions. This mechanism gives rise to flat, weakly dispersive bands in both Damon--Eshbach-like and backward-volume-like configurations, the latter of which has no direct counterpart in planar systems. Our results therefore establish a direct connection between planar chiral magnonic crystals and curvilinear nanomagnonics, and identify periodically engineered DMI in nanotubes as a route for controlling nonreciprocal spin waves, localized modes, and flat magnonic bands in three-dimensional magnetic architectures.

%
\begin{figure}[t]
\centering
\includegraphics[width=0.85\columnwidth]{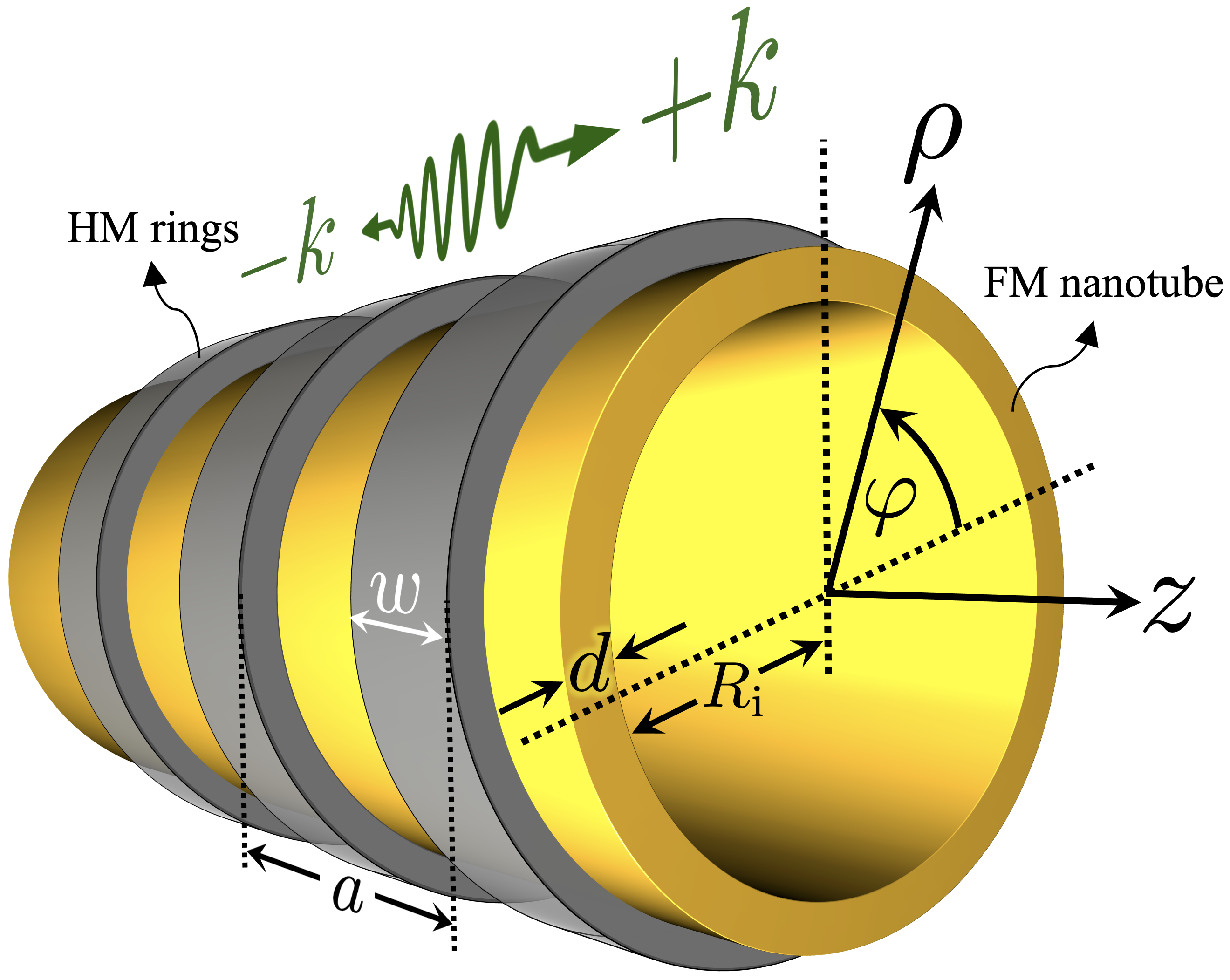}
\caption{Schematic representation of a chiral tubular magnonic crystal. A ferromagnetic nanotube of inner radius $R_{\rm i}$ and thickness $d$ is periodically covered by heavy-metal (HM) rings of width $w$, forming a one-dimensional axial lattice with period $a$. The HM rings induce an interfacial Dzyaloshinskii--Moriya interaction only in selected regions of the nanotube surface, giving rise to a periodic DMI profile $D(z)$. Spin waves propagate along the tube axis with wave vector $\mathbf{k}=\pm k\mathbf\,{\hat{z}}$. The cylindrical coordinate system $(\rho,\varphi,z)$ is also indicated.
}
\label{FIG1}
\end{figure}

\section{Theory: tubular magnonic crystal with periodic interfacial DMI}
\label{sec:theory_tube_dmi}

We consider a ferromagnetic nanotube of inner radius $R_{\rm i}$, outer radius $R$, and thickness $d=R-R_{\rm i}$. The tube axis is along the $z$ direction, and cylindrical coordinates $\mathbf r=\rho \hat{\boldsymbol\rho}+z\hat{\mathbf z}$ are used, with the local basis $\{\hat{\boldsymbol\rho},\hat{\boldsymbol\varphi},\hat{\mathbf z}\}$. 
The system is periodically modulated along the tube axis with period $a$. Physically, this modulation can be produced by covering the nanotube with heavy-metal rings, which induce an interfacial DMI only in selected axial regions, as shown in Fig.~\ref{FIG1}. In this way, the nanotube realizes a one-dimensional tubular magnonic crystal in which the spatially dependent DMI generates the periodic chiral potential.

Because we are interested in equilibrium configurations ranging from the axially saturated state to the vortex-like state, we consider a curling equilibrium magnetization lying in the local $\hat{\boldsymbol{\phi}}$--$\hat{\boldsymbol{z}}$ plane of the nanotube. Following the notation of Ref.~\cite{Mimica25}, the equilibrium magnetization direction is written as
\begin{equation}
\hat{\mathbf z}'=
\mathcal{C}\sin\theta\,\hat{\boldsymbol\varphi}
+
\cos\theta\,\hat{\mathbf z},
\label{eq:equilibrium_direction}
\end{equation}
so that $\mathbf M_0=M_{\rm s}\hat{\mathbf z}'$.
Here, $\theta$ measures the angle between the equilibrium magnetization and the tube axis, while $\mathcal{C}=\pm 1$ defines the chirality of the curling component. The limiting case $\theta=0$ corresponds to an axially saturated nanotube, whereas $\theta=\pi/2$ describes a flux-closure vortex state. In this work, we treat $\theta$ as a control parameter, enabling us to describe intermediate configurations between the vortex and saturated states.

The spin-wave dynamics are described in the local orthonormal basis $
\left\{
\hat{\boldsymbol\chi},
\hat{\boldsymbol\rho},
\hat{\mathbf z}'
\right\},
$
where
\begin{equation}
\hat{\boldsymbol\chi}
=
\hat{\boldsymbol\rho}\times\hat{\mathbf z}'
=
-\cos\theta\,\hat{\boldsymbol\varphi}
+
\mathcal{C}\sin\theta\,\hat{\mathbf z}.
\label{eq:chi_direction}
\end{equation}
Small-amplitude deviations around the equilibrium magnetization are then written as
\begin{equation}
\mathbf M(\mathbf r,t)
=M_{\rm s}\hat{\mathbf z}'
+ m_\chi(\mathbf r,t)\hat{\boldsymbol\chi}
+ m_\rho(\mathbf r,t)\hat{\boldsymbol\rho},
\label{eq:dynamic_magnetization_local}
\end{equation}
with $|m_\chi|,|m_\rho|\ll M_{\rm s}$.
In this notation, the cylindrical spin-wave amplitudes are expanded in the directions transverse to the equilibrium magnetization, while propagation occurs along the tube axis. The vortex and axially saturated configurations are therefore recovered as particular limits of the same formalism, and the intermediate curling state can be treated without changing the structure of the eigenvalue problem.

The magnetization dynamics is governed by the Landau--Lifshitz equation
\begin{equation}
\frac{\partial \mathbf M}{\partial t}
=
-\gamma\mu_0 \mathbf M\times \mathbf H_{\rm eff},
\label{eq:LL_tube}
\end{equation}
where $\gamma$ is the absolute value of the gyromagnetic ratio and $\mathbf H_{\rm eff}$ is the effective magnetic field. In the present problem,
\begin{equation}
\mathbf H_{\rm eff}
=
\mathbf H_0
+
\mathbf H_{\rm u}
+
\mathbf H_{\rm ex}
+
\mathbf H_{\rm dip}
+
\mathbf H_{\rm dmi}.
\label{eq:Heff_tube}
\end{equation}
Here, $\mathbf H_0$ is the applied magnetic field, $\mathbf H_{\rm u}$ is the uniaxial anisotropy field, $\mathbf H_{\rm ex}$ is the exchange field, $\mathbf H_{\rm dip}$ is the magnetostatic field, and $\mathbf H_{\rm dmi}$ is the interfacial DMI field induced by the heavy-metal coating.

The key ingredient of the tubular magnonic crystal is the periodic axial modulation of the DMI constant, $D(z+a)=D(z)$.In the present model, we isolate the effect of the periodic DMI and neglect
possible concomitant modulations of other magnetic parameters induced by
the heavy-metal patterning.
The periodic DM constant $D(z)$ can be expanded as
\begin{equation}
D(z)=\sum_{n} D_{G_n} e^{iG_{n}z},
\qquad
G_{n}=\frac{2\pi n}{a},
\label{eq:D_Fourier_tube}
\end{equation}
where $n$ is an integer. The Fourier coefficients are
\begin{equation}
D_{G_n}=
\frac{1}{a}
\int_{-a/2}^{a/2} D(z)e^{-iG_n z}\,dz.
\label{eq:DG_general}
\end{equation}
For a binary axial modulation, $D(z)=D$ within a region of width $w$, centered in each unit cell, and $D(z)=0$ in the remaining part of the unit cell, one obtains
\begin{equation}
D_{G_n} = \frac{D}{a}
w
\operatorname{sinc}
\left(
\frac{G_n w}{2}
\right),
\label{eq:DG_rings}
\end{equation}
where $w$ is the width of each HM ring (DMI-active region), and $\operatorname{sinc}(x)=\sin x/x$.
Because the modulation is periodic along $z$, the dynamic magnetization follows Bloch's theorem. 
 
For a given axial component of the wave vector $k$ in the first Brillouin zone and azimuthal index $\ell$, the dynamic magnetization components are expanded in the local transverse basis as
\begin{equation}
m_\nu(\varphi,z,t)
=
\sum_{G_n}
m_{\nu,G_n}
e^{i[(k+G_n)z+\ell\varphi-\omega t]},
\label{eq:Bloch_tube}
\end{equation}
where $\nu=\chi,\rho$. The components $m_\chi$ and $m_\rho$ describe oscillations perpendicular to the equilibrium magnetization $\mathbf M_0=M_{\rm s}\hat{\mathbf z}'$. In the thin-shell approximation, the lowest radial mode is taken as uniform across the tube thickness, so that the radial dependence is averaged out and the dynamics is described by the amplitudes $m_{\nu,G_n}$. Substitution of Eq.~\eqref{eq:Bloch_tube} into the linearized Landau--Lifshitz equation leads to an eigenvalue problem
\begin{equation}
\omega
\begin{pmatrix}
m_{\chi,G_n} \\
m_{\rho,G_n}
\end{pmatrix}
=
i\gamma\mu_0 M_{\rm s}
\sum_{n'}
\begin{pmatrix}
T_{\chi\chi}^{nn'} & T_{\chi\rho}^{nn'} \\
T_{\rho\chi}^{nn'} & T_{\rho\rho}^{nn'}
\end{pmatrix}
\begin{pmatrix}
m_{\chi,G_{n'}} \\
m_{\rho,G_{n'}}
\end{pmatrix}.
\label{eq:eigen_tube}
\end{equation}
Equivalently,
\begin{equation}
\omega \mathbf m
=
i\gamma\mu_0 M_{\rm s} \mathcal{T}(k)\mathbf m ,
\label{eq:eigen_compact_tube}
\end{equation}
where $\mathcal{T}(k)$ is the dynamical matrix of the tubular magnonic crystal. Appendix~\ref{AppAA} derives the effective field associated with the spatially periodic DMI. The complete matrix elements of $\mathcal{T}(k)$, incorporating the DMI field together with the remaining magnetic interactions, are given in Appendix~\ref{AppBB}. Diagonalization of $\mathcal{T}(k)$ provides the eigenfrequencies $\omega_j(k)$ and their corresponding eigenvectors, with the bands $f_j(k)=\omega_j(k)/(2\pi)$ defining the spin-wave spectrum.
\\


\section{Results and discussion}
\label{sec:results}

\begin{figure*}[t]
\includegraphics[width=1\textwidth]{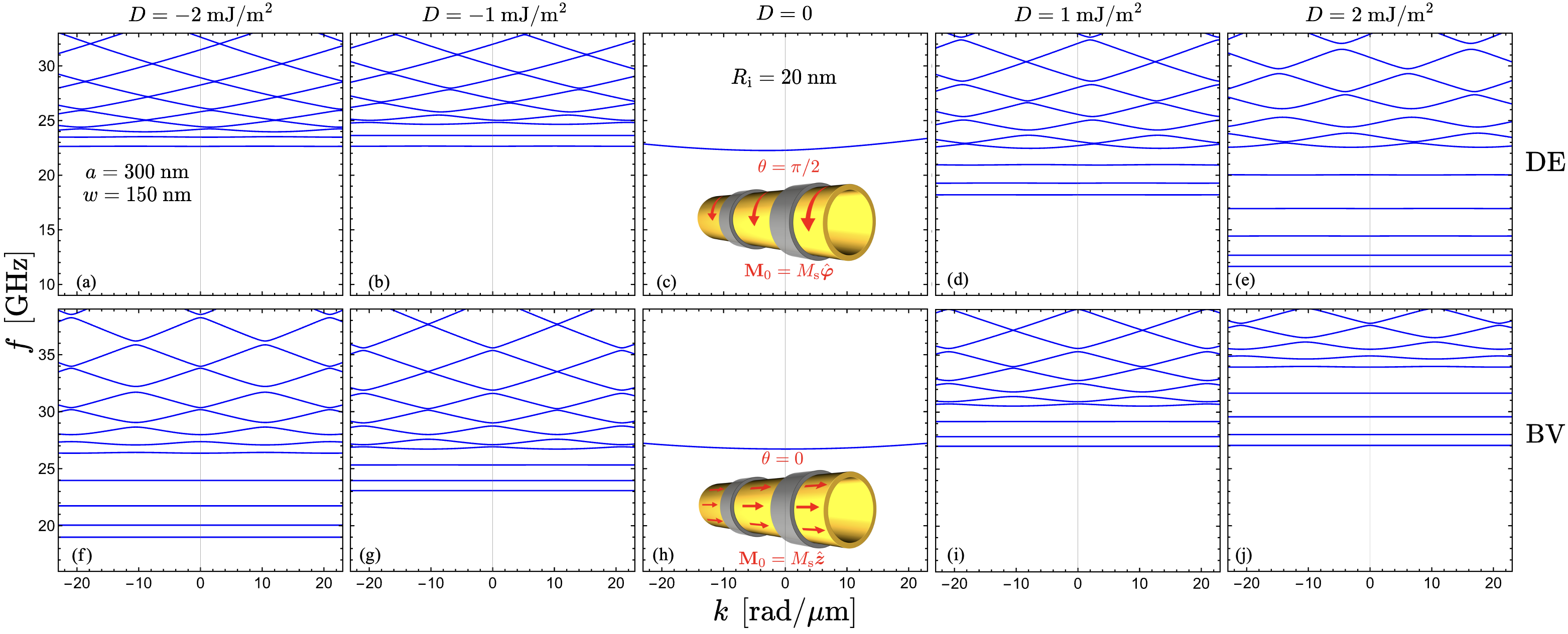}
\caption{
Spin-wave band structures of the chiral tubular magnonic crystal evaluated at $\mu_0H_0=500$~mT and for different values of the interfacial Dzyaloshinskii--Moriya constant $D$. The bands are shown for $D=-2,-1,0,1$, and $2~\mathrm{mJ/m^2}$. The upper and lower rows correspond to the Damon--Eshbach-like (a--e) and backward-volume-like (f--j) configurations, respectively. The calculations are performed for a periodic DMI modulation with lattice constant $a=300~\mathrm{nm}$ and DMI-active ring width $w=150~\mathrm{nm}$. The nanotube has inner radius $R_{\rm i}=20~\mathrm{nm}$ and thickness $d=1~\mathrm{nm}$. The vertical dashed lines indicate the Brillouin edges associated with the axial modulation. For $D=0$, the uncoupled folded replicas are omitted, and the unfolded dispersion of the uniform nanotube is shown for reference.}
  \label{FIG2}
\end{figure*}

In the following, we apply the plane-wave formalism developed above to a one-dimensional tubular magnonic crystal based on Permalloy ($\mathrm{Ni}_{80}\mathrm{Fe}_{20}$).Unless otherwise stated, we use $M_s=796~\mathrm{kA/m}$,
$A_{\rm ex}=13~\mathrm{pJ/m}$, $\gamma/2\pi=28.0~\mathrm{GHz/T}$,
$\mu_0H_0=500~\mathrm{mT}$, $H_u=0$, and a nanotube thickness
$d=1~\mathrm{nm}$. The magnonic crystal is formed by periodically patterning heavy-metal rings along the tube axis, so that the interfacial Dzyaloshinskii--Moriya interaction is induced only in the FM regions in contact with the rings. This produces a periodic DMI profile, $D(z)$, with lattice constant $a$, as shown in Fig.~\ref{FIG1}.
For completeness, the formalism developed above retains an arbitrary azimuthal index $\ell$ and chirality $\mathcal{C}$. The numerical calculations and discussion presented below, however, are restricted to the azimuthally uniform sector $\ell=0$ and $\mathcal{C}=1$.
In the numerical calculations, the plane-wave expansion is truncated to the $2N+1$ reciprocal lattice vectors $G_n$, with $n=-N,\ldots, N$ and $|G_n|\leq G_{\rm N}$. The resulting dynamical matrix has dimension $2(2N+1)\times 2(2N+1)$. We determine the cutoff $N$ through a convergence analysis by increasing the basis size until the low-frequency modes of interest remain essentially unchanged. In our case, we use $N=40$.

We evaluate the spin-wave spectrum for nanotubes with tangential equilibrium states. To stabilize these configurations, we apply a bias field along the local equilibrium direction, $\mathbf{H}_0=H_0\hat{\mathbf{z}}'$, where $\hat{\mathbf{z}}'$ is defined in Eq.~\eqref{eq:equilibrium_direction}. This can be achieved with a combination of an axial field and a circumferential field. Strictly speaking, the spatial variation of $D(z)$ generates localized static DMI fields at the boundaries between DMI-active and DMI-free regions, which may induce a small local canting of the equilibrium magnetization in configurations with a finite axial component. Here, we neglect this canting and assume that the applied bias field is sufficiently strong to maintain the prescribed equilibrium direction $\hat{\mathbf{z}}'$ throughout the unit cell.


\begin{figure*}[t]
\includegraphics[width=1\textwidth]{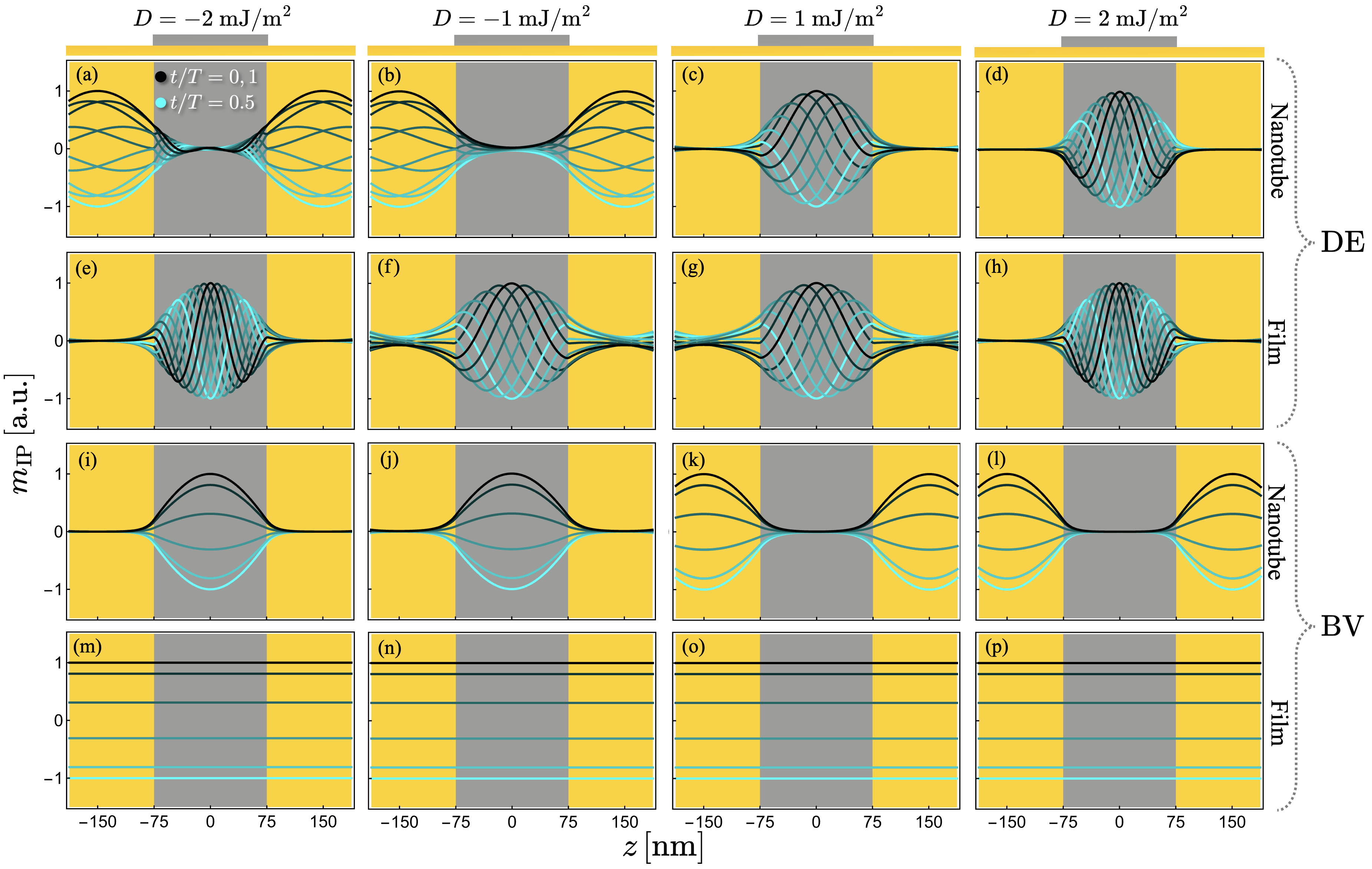}
\caption{
Spatial profiles of the lowest-frequency spin-wave eigenmodes in the chiral tubular magnonic crystal for different values of the interfacial Dzyaloshinskii--Moriya constant. The profiles are calculated using the in-plane dynamic magnetization component $m_{\rm IP}=m_{\chi}$ evaluated at $k=0$. The background colors indicate the DMI-active/HM-covered regions (gray) and the DMI-free regions (yellow). The columns correspond to $D=-2$, $-1$, $1$, and $2~\mathrm{mJ/m^2}$. The upper blocks show the Damon--Eshbach-like configuration, with the nanotube profiles in panels (a--d) and the corresponding planar-film profiles in panels (e--h). The lower blocks show the backward-volume-like configuration, with the nanotube profiles in panels (i--l) and the planar-film profiles in panels (m--p). The plotted quantity is the normalized dynamic magnetization component $m_{\rm IP}$ as a function of the axial coordinate $z$. The curves correspond to different phases within one oscillation period. The vertical dashed lines indicate the boundaries between DMI-active and DMI-free regions of the unit cell, as indicated by the upper bars.
}
  \label{FIG3}
\end{figure*}

With the geometry and magnetic background established, we now analyze how the periodic DMI reshapes the spin-wave spectrum, focusing in particular on the emergence of band gaps, weakly dispersive bands, and localized modes associated with the DMI-active and DMI-free regions of the unit cell.
We first examine the effect of the magnitude and sign of the interfacial DMI on the spin-wave band structure. Figure~\ref{FIG2} shows the band structure of the tubular magnonic crystal for different values of the DMI constant, $D=-2,-1,0,1$, and $2~\mathrm{mJ/m^2}$. The upper and lower rows correspond, respectively, to the vortex state ($\theta=\pi/2,\mathbf{k}\perp\mathbf{M}_0$) and the axial state ($\theta=0,\mathbf{k}\parallel\mathbf{M}_0$), hereafter referred to as the Damon–Eshbach-like (DE) and backward-volume-like (BV) configurations. As shown in Figs.~\ref{FIG2}(c,h), in the absence of DMI, the axial modulation is effectively removed, and the spectrum corresponds to that of a uniform nanotube without band gaps. Once the periodic interfacial DMI is introduced, it strongly modifies the band structure. The DMI-active regions act as a chiral axial modulation, coupling different reciprocal-lattice components and producing avoided crossings, band gaps, and nearly dispersionless bands.

A central feature of Fig.~\ref{FIG2} is that flat and weakly dispersive bands appear in both DE and BV configurations. This result is particularly relevant because it has no direct analogy in planar 1D chiral magnonic crystals. In planar films with periodic interfacial DMI, the strongest modulation of bands is expected in the DE geometry, where the DMI contribution produces a linear-in-$k$ nonreciprocal shift \cite{Gallardo19c}. In contrast, for the BV geometry, where $\mathbf{k}\parallel \mathbf{M}_0$, the DMI contribution vanishes, and no flat bands are expected. The tubular case differs qualitatively from this planar scenario. Here, the cylindrical geometry combines axial propagation, curvature, and the curling character of the equilibrium magnetization. As a result, the DMI matrix elements shown in Appendix~\ref{AppBB} depend not only on the axial wave vector but also on the azimuthal index, the equilibrium angle, and curvature-dependent terms.

The spectra also display a pronounced dependence on the sign of $D$. In planar chiral magnonic crystals, reversing the DMI constant mainly reverses the odd-in-$k$ frequency shift, while leaving the overall gap and localization patterns essentially unchanged \cite{Gallardo19c,Flores22}. In the tubular geometry, however, changing $D\rightarrow-D$ also reverses the curvature-induced DMI contribution to the diagonal effective field, as discussed below in connection with Eq.~\eqref{Hueff}. Consequently, the sign of $D$ modifies not only the direction of the frequency nonreciprocity, but also the local frequency contrast between the DMI-active and DMI-free segments. This results in different band shifts, gap openings, and distributions of weakly dispersive bands for positive and negative DMI.

The effect of increasing $|D|$ can therefore be stated more precisely. For $|D|=1~\mathrm{mJ/m^2}$, the spectra already exhibit DMI-induced gaps and weakly dispersive branches. At $|D|=2~\mathrm{mJ/m^2}$, the larger contrast between the DMI-active and DMI-free regions produces additional anticrossings and a larger number of nearly flat branches. Their frequency evolution, however, depends jointly on the sign of $D$ and the magnetic configuration. In the DE configuration, the downward displacement of the low-frequency flat bands is stronger for $D>0$ [Figs.~\ref{FIG2}(d,e)], whereas for $D<0$ [Figs.~\ref{FIG2}(a,b)] several branches are slightly shifted down and remain comparatively close to the frequency scale of the DMI-free regions. In the BV configuration, this trend is reversed: the pronounced downward shift occurs for $D<0$ [Figs.~\ref{FIG2}(f,g)], while for $D>0$ [Figs.~\ref{FIG2}(i,j)] the lowest weakly dispersive branches remain comparatively pinned.

This sign-dependent evolution of $D$ anticipates the effective-field interpretation developed below. When the curvature-induced DMI contribution lowers the local frequency of the DMI-covered segments, these regions act as potential wells and host localized modes. For the opposite sign of $D$, the covered segments behave as barriers, and the lowest-frequency modes are displaced toward the DMI-free portions of the unit cell, whose local frequency is not directly shifted by the DMI.
Within the convention adopted here, reversing the interface normal (for example, by moving a given HM/FM interface from the outer to the inner surface of the nanotube) reverses the corresponding sign of $D$. For a fixed outer coating, however, both the magnitude and sign of the interfacial DMI also depend on the specific HM/FM material combination \cite{Kuepferling23}. The tubular magnonic crystal is therefore sensitive to both the interface orientation and the material-dependent DMI.

Overall, Fig.~\ref{FIG2} demonstrates that a periodically modulated interfacial DMI produces a tubular band structure that is not a straightforward extension of its planar counterpart. The simultaneous action of Bragg scattering, interfacial chirality, curvature, azimuthal boundary conditions, and the curling equilibrium state gives rise to nearly flat bands in both DE and BV configurations. 
The spatial character of these modes and their relation to the DMI-active and DMI-free regions are examined below through the corresponding eigenmode profiles.

To gain further insight into the physical origin of the flat and weakly dispersive bands discussed above, Fig.~\ref{FIG3} displays the corresponding spin-wave mode profiles. The profiles are evaluated for $D=-2,-1,1$, and $2~\mathrm{mJ/m^2}$ and correspond to the lowest-frequency eigenmode of each spectrum. For the nearly flat bands, the spatial distribution depends weakly on the wave vector $k$; therefore, the specific $k$ value used to plot the profile does not modify the qualitative localization pattern. For comparison, we also show the corresponding eigenmodes of an equivalent planar film with the same periodic DMI modulation.

The contrast between planar and tubular geometries is immediately evident, as shown in Figs.~\ref{FIG3}(a--h) for the DE configuration. In the planar film [see Figs.~\ref{FIG3}(e--h)], reversing the sign of the DMI constant mainly reverses the nonreciprocal frequency shift, while the spatial localization of the low-frequency modes remains essentially unchanged. In the nanotube considered in Figs.~\ref{FIG3}(a--d), however, the situation is fundamentally different: changing $D\rightarrow -D$ modifies not only the dispersion, as already observed in Fig.~\ref{FIG2}, but also the spatial character of the eigenmodes. Thus, in the tubular geometry, the localization of the flat-band modes depends not only on the magnitude of the DMI but also on its sign.

The difference is even more pronounced in the BV configuration shown in Figs.~\ref{FIG3}(i--p). In the planar film illustrated in Figs.~\ref{FIG3}(m--p), the interfacial DMI produces no appreciable effect because the spin waves propagate parallel to the equilibrium magnetization. In this geometry, the conventional DMI contribution to the spin-wave dispersion vanishes \cite{Cortes13}, so that the film profiles remain nearly uniform and essentially insensitive to the sign of $D$. In the nanotube [see Figs.~\ref{FIG3}(i--l)], by contrast, curvature changes the way the interfacial DMI enters the magnetization dynamics, allowing it to remain effective even in the BV configuration. The periodically modulated DMI therefore induces an axial modulation of the local magnetic environment, creating regions where the spin-wave amplitude becomes strongly localized. This curvature-enabled localization mechanism has no direct counterpart in planar chiral magnonic crystals.


\begin{figure*}[t]
\includegraphics[width=1\textwidth]{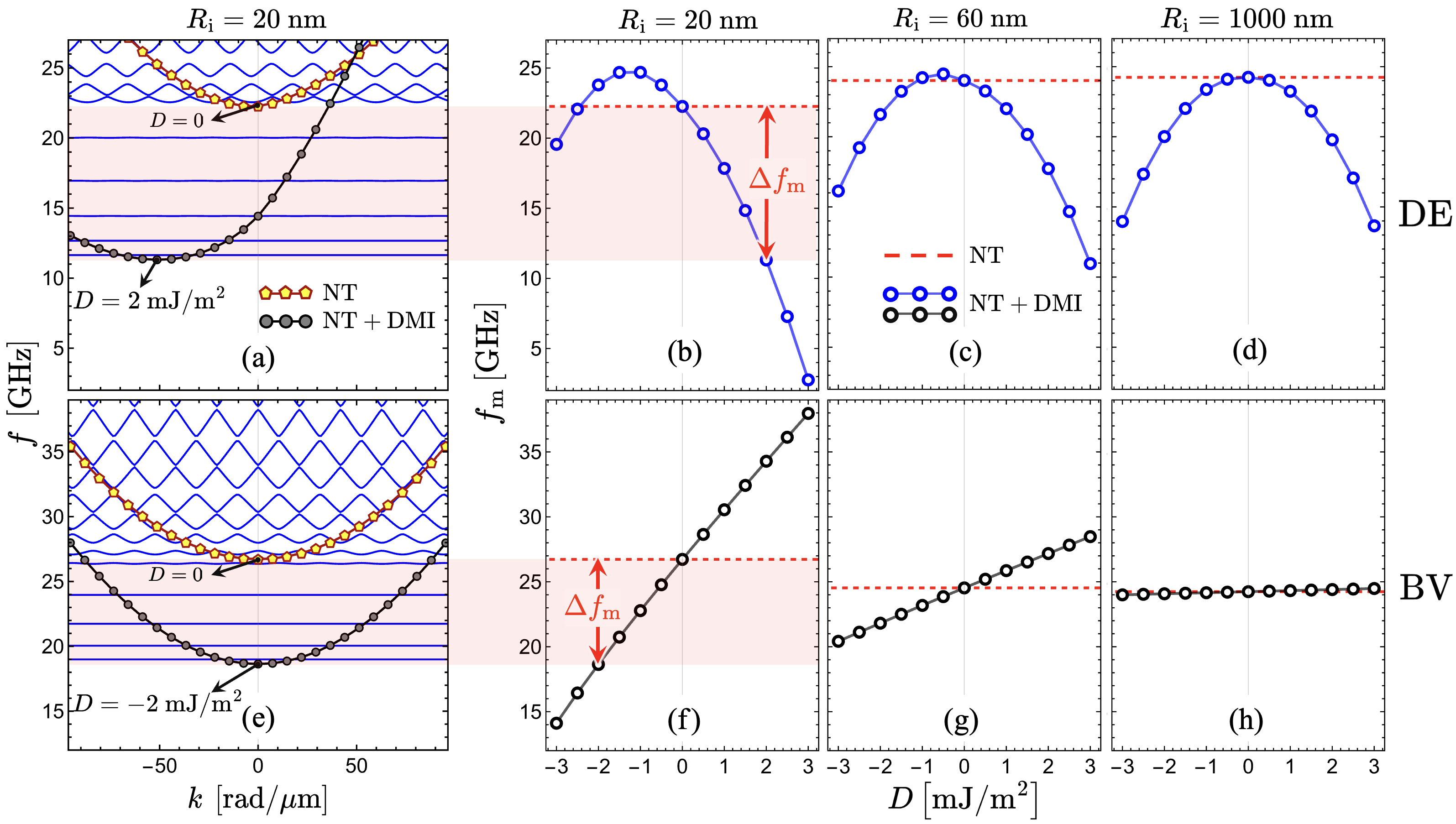}
\caption{Connection between the band structure of the periodically modulated nanotube and the continuous-nanotube limit with and without uniform DMI. Panels (a) and (e) compare the bands of the periodic tubular magnonic crystal (blue curves) with the dispersions of a continuous nanotube without DMI (yellow diamonds) and with uniform DMI (gray circles), for $R_i=20~\mathrm{nm}$. Panel (a) corresponds to the DE configuration with $D=2~\mathrm{mJ/m^2}$, whereas panel (e) corresponds to the BV configuration with $D=-2~\mathrm{mJ/m^2}$. Panels (b–d) and (f–h) show the minimum frequency $f_m$ of the continuous nanotube as a function of $D$ for $R_i=20$, 60, and $1000~\mathrm{nm}$, in the DE and BV configurations, respectively. The dashed horizontal lines indicate the corresponding DMI-free values. In the DE configuration, the sign-dependent asymmetry observed at small radii is progressively suppressed as the radius increases, recovering the nearly sign-symmetric planar-film behavior. In the BV configuration, the DMI-induced frequency shift is entirely curvature-driven and therefore vanishes in the large-radius limit. The intervals $\Delta f_m$ indicated in panels (b) and (f), evaluated at $D=2~\mathrm{mJ/m^2}$ and $D=-2~\mathrm{mJ/m^2}$, respectively, are bounded by the minimum frequencies of the DMI-free and uniformly DMI-covered nanotubes. These intervals estimate the frequency windows in which DMI-induced flat bands are expected in the periodically modulated system.}
  \label{FIG4}
\end{figure*}

To further clarify the role of curvature and the sign of the interfacial DMI in the tubular chiral magnonic crystal, we analyze the limiting case of a continuous nanotube with uniform DMI.
This limit is obtained from the analytical expressions derived in Appendix~\ref{AppBB} by setting $G_n=0$ and $w=a$, thereby removing the axial modulation while preserving the cylindrical geometry. In this case, the spin-wave frequency agrees with previous works \cite{Mimica25,Mimica26} and is given by 
\begin{widetext}
\begin{equation}
f
=\!
\frac{\gamma \mu_0M_{\rm s}}{2\pi}
\left\{i\mathcal{N}_{\chi\rho}^{\rm dip}+\frac{2Dk}{\mu_0M_{\rm s}^2}
\mathcal{C}\sin\theta
+\!
\sqrt{
\left(
\frac{H_0}{M_{\rm s}}
+
\frac{H_{\rm u}^{\rm eff}}{M_\text{s}}\cos^2\theta+k^2\lambda_\text{ex}^2+\mathcal{N}_{\rho\rho}^{\rm dip}
\right)\!\!
\left(
\frac{H_0}{M_{\rm s}}
+
\frac{H_{\rm u}^{\rm eff}}{M_\text{s}}\cos2\theta+k^2\lambda_\text{ex}^2+\mathcal{N}_{\chi\chi}^{\rm dip}
\right)
}
\right\}.
\label{fan}
\end{equation}
\end{widetext}
Here, we consider the case $\ell=0$ (azimuthally uniform mode) and we have defined an effective anisotropy field 
\begin{equation}
H_{\rm u}^{\rm eff}
=
H_{\rm u}
+
\frac{2\pi M_{\rm s}\lambda_{\rm ex}^2}{S}
\ln\!\left(\!\frac{R}{R_{\rm i}}\!\right)
+
\frac{4D}
{\mu_0 M_{\rm s}\!\left(R+R_{\rm i}\right)},
\label{Hueff}
\end{equation}
where $\lambda_{\rm ex}=\sqrt{2A_{\rm ex}/\mu_0M_{\rm s}^2}$ is the exchange length, $S=\pi(R^2-R_{\rm i}^2)$ is the nanotube cross-section and the quantities $\mathcal{N}_{\chi\rho}^{\rm dip}$, $\mathcal{N}_{\rho\rho}^{\rm dip}$, and $\mathcal{N}_{\chi\chi}^{\rm dip}$ are the dipolar matrix elements given by
\begin{equation}
    \mathcal{N}^{\rm dip}_{\chi\rho}=-\frac{ik \, \mathcal{C}\sin\theta}{S}\!\!\int_{R_{\rm i}}^{R}\!\partial_\rho\!\left[\int_{R_{\rm i}}^{R}\!\mathcal{G}_{0}(\rho,\rho')\rho'd\rho'\right]\!\rho 
    d\rho,
\end{equation}
\begin{equation}
    \mathcal{N}^{\rm dip}_{\rho\rho}=\!\frac{1}{S}\!\int_{R_{\rm i}}^{R}\!\!\partial_\rho\!
    \left[
       \mathcal{G}_{0}(\rho,\rho')\rho'\vert ^{\rho'=R}_{\rho'=R_{\rm i}}
    -\!
 \int_{R_{\rm i}}^{R}\! \!\mathcal{G}_{0}(\rho,\rho') d\rho'\!
    \right]\!\rho d\rho,
\end{equation}
and 
\begin{equation}
    \mathcal{N}_{\chi\chi}^{\rm dip}=\frac{k^2\sin^2\theta}{S}\int_{R_{\rm i}}^{R}\left[\int_{R_{\rm i}}^{R} \mathcal{G}_{0}(\rho,\rho')\rho'd\rho'\right]\rho d\rho.
\end{equation}
The function $\mathcal{G}_{0}(\rho,\rho')$ is defined as 
\begin{equation}
     \mathcal{G}_{0}(\rho,\rho') = 2\pi\, I_{0}\bigl(|k|\,\rho_<\bigr)\, K_{0}\bigl(|k|\,\rho_>\bigr),
\end{equation}
where $I_{0}(x)$ and $K_{0}(x)$ denote, respectively, the modified Bessel functions of the first and second kind of zero order. Furthermore, the variables $\rho_<$ and $\rho_>$ are given by $\rho_< = \min(\rho,\rho')$, and $\rho_> = \max(\rho,\rho')$.

Equation~\eqref{fan} separates two different roles of the interfacial DMI. The term proportional to $Dk\sin\theta$ is the conventional interfacial DMI contribution found in planar FM films coupled with heavy metals \cite{Tacchi17}. It is odd in the wave vector and causes the usual DMI-induced frequency nonreciprocity. By contrast, the DMI contribution contained in $H_{\rm u}^{\rm eff}$ is induced by curvature. It scales as $(R+R_{\rm i})^{-1}$ and therefore vanishes in the planar limit $R_{\rm i}\rightarrow\infty$.

Equation~\eqref{Hueff} shows that, for the azimuthally uniform mode, both the exchange-driven and DMI-driven curvature corrections enter the dispersion as an effective uniaxial anisotropy field.  This result is consistent with the general framework of curvilinear micromagnetism, in which the exchange and DMI energies are decomposed into conventional and geometry-induced contributions with anisotropy- and DMI-like symmetries \cite{Sheka22,Sheka22b}. In particular, interfacial DMI in a curved shell generates an emergent uniaxial anisotropy-like term proportional to the product of the DMI strength and the mean curvature. For the cylindrical shell considered here, the corresponding anisotropy field gives precisely the DMI-dependent term in Eq.~\eqref{Hueff}. 
This provides a simple local interpretation of the periodic tubular magnonic crystal. In the DMI-active regions, $D\neq0$ and therefore $H_{\rm u}^{\rm eff}$ is shifted by this curvature-dependent term, whereas in the uncovered regions, $D=0$ and the shift is absent. Thus, the periodic DMI acts not only as a spatially modulated chiral interaction, but also as a periodic modulation of an effective anisotropy field enabled by curvature. 
Depending on the sign of $D$ and the magnetic configuration, this local field may either reduce or increase the spin-wave frequency, thereby determining whether the DMI-covered regions act as wells or barriers for the lowest-frequency modes.

Let us first consider the DE configuration for which $\theta=\pi/2$. In this case, for $\mathcal{C}=1$, Eq.~\eqref{fan} reduces to
\begin{eqnarray}
f_{\rm DE}\!
&=&\!
\frac{\gamma\mu_0M_{\rm s}}{2\pi}
\left\{
i\mathcal{N}_{\chi\rho}^{\rm dip}
+
\frac{2Dk}{\mu_0M_{\rm s}^2}
\right.\notag
\\
&&\!\!\!\!+\!
\left.
\sqrt{
\!\left(\!
\frac{\mathcal{B}_k}{M_{\rm s}}+\mathcal{N}_{\rho\rho}^{\rm dip}
\!\right)\!\!
\left(
\!\frac{\mathcal{B}_k}{M_{\rm s}}
-
\frac{H_{\rm u}^{\rm eff}}{M_{\rm s}}+\mathcal{N}_{\chi\chi}^{\rm dip}
\!
\right)
}
\right\},
\label{fDE}
\end{eqnarray}
where $\mathcal{B}_k=H_0+M_{\rm s}\lambda_\text{ex}^2 k^2$. The first DMI term, proportional to $Dk$, shifts the dispersion in opposite directions for opposite signs of $D$. Therefore, this term mainly reverses the direction of the nonreciprocity when $D$ changes sign, but it does not invert the localization landscape. The curvature-induced DMI contribution contained in $H_{\rm u}^{\rm eff}$ has a different effect. Since it enters the DE-like frequency with a minus sign, a positive DMI increases $H_{\rm u}^{\rm eff}$ and lowers the local spin-wave frequency in the DMI-active regions. These regions then behave as effective potential wells for the low-frequency modes. Conversely, a negative DMI reduces $H_{\rm u}^{\rm eff}$ in the DMI-active regions, increasing their local frequency relative to the uncovered parts of the unit cell. In that case, the DMI-free regions become the effective wells. This explains the sign-dependent localization observed in the nanotube profiles shown in Figs.~\ref{FIG3}(a--d). In the corresponding planar film, Figs.~\ref{FIG3}(e--h), the curvature-dependent contribution vanishes, and only the usual $Dk$ term remains, so the profiles are essentially insensitive to the sign of $D$.

We now turn to the BV-like configuration, corresponding to $\theta=0$. In this case, 
 the linear-in-$k$ DMI contribution vanishes, and Eq.~\eqref{fan} becomes
%
%
\begin{equation}
f_{\rm BV}
\!=\!
\frac{\gamma\mu_0M_{\rm s}}{2\pi}
\!\sqrt{
\!\left(
\!\frac{\mathcal{B}_k}{M_{\rm s}}
\!+\!
\frac{H_{\rm u}^{\rm eff}}{M_{\rm s}}\!+\!\mathcal{N}_{\rho\rho}^{\rm dip}
\!\right)\!\!
\left(\!
\frac{\mathcal{B}_k}{M_{\rm s}}
\!+\!
\frac{H_{\rm u}^{\rm eff}}{M_{\rm s}}\!+\!\mathcal{N}_{\chi\chi}^{\rm dip}
\!\right)
 }.
\end{equation}
This expression clarifies why the BV-like response of the tubular system has no planar counterpart. In a planar film, the term $Dk\sin\theta$ vanishes and the curvature-dependent DMI contribution in $H_{\rm u}^{\rm eff}$ also disappears as $R_{\rm i}\rightarrow\infty$. Therefore, the BV frequency becomes independent of the DMI constant. In a nanotube, however, the DMI contribution to $H_{\rm u}^{\rm eff}$ remains finite and shifts both diagonal effective-field terms. For $D<0$, this shift reduces $H_{\rm u}^{\rm eff}$ and lowers the local spin-wave frequency in the DMI-active regions, which then host the low-frequency localized modes. For $D>0$, the local frequency is increased in the DMI-active regions, and the lowest-frequency modes are expelled toward the DMI-free portions of the unit cell. This is consistent with the nanotube profiles shown in Figs.~\ref{FIG3}(i--l), whereas the planar profiles in Figs.~\ref{FIG3}(m--p) remain nearly uniform.

Figure \ref{FIG4} connects the flat bands of the periodically modulated nanotube with the dispersion of a continuous nanotube possessing uniform DMI and quantifies the dependence of this mechanism on curvature. Panels \ref{FIG4}(a) and \ref{FIG4}(e), for DE and BV configurations, show representative cases in which the DMI-covered regions act as local frequency wells. In the DE configuration, Fig.~\ref{FIG4}(a), a positive DMI shifts the continuous-nanotube dispersion along the wave-vector axis through the conventional odd-in-$k$ term and simultaneously lowers its minimum through the curvature-induced DMI contribution to $H_{\rm u}^{\rm eff}$. The resulting reduction relative to the DMI-free nanotube is denoted by $\Delta f_{\rm m}$, as defined in Figs.~\ref{FIG4}(b,f). When the DMI is periodically modulated, the alternating local dispersions hybridize, producing a set of low-frequency, weakly dispersive bands shown by the blue curves in Figs.~\ref{FIG4}(a,e).

In the BV configuration, Fig.~\ref{FIG4}(e), the situation is qualitatively different. Because $Dk\sin\theta=0$ in Eq. \eqref{fan} for $\theta=0$, the uniformly DMI-covered nanotube remains reciprocal and its minimum stays at $k=0$. Nevertheless, a negative DMI lowers the entire low-frequency branch through the curvature-induced contribution to $H_{\rm u}^{\mathrm{eff}}$. The periodic alternation between this reduced local frequency and that of the DMI-free regions then produces the nearly flat BV bands. Thus, panels \ref{FIG4}(a) and \ref{FIG4}(e) directly illustrate the different origins of the frequency reduction in the two configurations: both the conventional and curvature-induced DMI terms contribute in DE, whereas only the curvature-induced term operates in BV.

Figures \ref{FIG4}(b–d), for the DE configuration, show the minimum frequency as a function of $D$. For small radii, the combination of the conventional nonreciprocal shift and the curvature-induced effective-field contribution produces a strongly asymmetric and nonmonotonic dependence on $D$. As the radius increases, the DMI-dependent contribution to $H_{\rm u}^{\mathrm{eff}}$, namely $4D/[\mu_0M_{\rm s}(R+R_{\rm i})]$, progressively vanishes. The remaining planar contribution is odd in $k$, so reversing the sign of $D$ approximately mirrors the dispersion about $k=0$ without changing its minimum frequency. Consequently, $f_{\rm m}(D)$ approaches a nearly even function of $D$ in the large-radius limit.
The curvature origin of the effect is even clearer in the BV configuration, as shown in Figs.~\ref{FIG4}(f–h). Since the conventional $Dk\sin\theta$ term vanishes in this configuration, any dependence on $D$ arises from the curvature-induced contribution to $H_{\rm u}^{\mathrm{eff}}$. For small radii, negative DMI reduces $H_{\rm u}^{\mathrm{eff}}$ and lowers $f_{\rm m}$, whereas positive DMI increases both quantities. The slope of $f_{\rm m}(D)$ decreases rapidly with increasing radius because the anisotropy-like DMI contribution scales as $(R+R_{\rm i})^{-1}$. For $R_{\rm i}=1000~\mathrm{nm}$, the minimum frequency is therefore nearly independent of $D$, as expected in the planar BV limit (see dashed red line).

Overall, this effective-anisotropy picture provides a compact interpretation of the flat bands and localized modes in the chiral tubular magnonic crystal. The periodic DMI generates a local frequency landscape through the curvature-induced contribution to $H_{\rm u}^{\rm eff}$, while the axial periodicity provides the Bragg scattering required to form magnonic bands. The sign of $D$ determines whether the DMI-covered regions behave as potential wells or barriers, and the disappearance of this effect for large radii confirms its curvature-enabled origin.

\section{Conclusions}

We have introduced a chiral tubular magnonic crystal formed by a ferromagnetic nanotube with a periodically modulated interfacial Dzyaloshinskii--Moriya interaction. The modulation, which may be realized by periodically covering the nanotube with heavy-metal rings, produces alternating DMI-active and DMI-free regions along the tube axis. We developed a plane-wave formalism adapted to the cylindrical geometry, allowing us to calculate the spin-wave band structure while retaining exchange, dipolar interactions, curvature, and interfacial DMI.

Our results show that the tubular geometry qualitatively modifies the role of periodic DMI compared with planar chiral magnonic crystals. Besides the usual nonreciprocal term proportional to $Dk$, the curvature converts part of the interfacial DMI into an anisotropy-like effective internal field \cite{Sheka22}. As a consequence, the DMI sign controls the local frequency landscape within the unit cell: DMI-covered regions may act as potential wells or as barriers for the lowest-frequency modes. This mechanism explains the sign-dependent localization observed in the nanotube and the appearance of flat and weakly dispersive bands in both Damon--Eshbach-like and backward-volume-like configurations.

The analysis of the continuous nanotube limit further confirms that curvature drives this effect. For small radii, the minimum frequency depends strongly and asymmetrically on the sign of the DMI, whereas in the large-radius limit the curvature-induced contribution vanishes and the standard planar behavior is recovered. In this limit, the backward-volume response becomes independent of DMI. These findings establish that periodically engineered interfacial DMI in nanotubes provides a route to controlling nonreciprocal spin waves, localized modes, and flat magnonic bands in curved magnetic architectures.
\\
\section*{ACKNOWLEDGMENTS}

We acknowledge financial support from FONDECYT, grants 1250803 and 1241589,  and CEDENNA under grant CIA250002 from ANID. J.F.-F. acknowledges support from ANID FONDECYT Postdoctoral Grant 3240558.

\appendix

\renewcommand{\thesubsection}{\Alph{section}\arabic{subsection}}

\section{Periodic Dzyaloshinskii-Moriya field}
\label{AppAA}
To incorporate the spatial dependence of the interfacial DMI, we calculate the corresponding effective field as follows. The Dzyaloshinskii–Moriya Hamiltonian is given by \cite{Fert80,Crepieux98}
\begin{equation}
\mathcal{H}^{\text{DM}}=\sum_{i}\mathbf{D}_{i-1,i}\cdot(\mathbf{S}_{i-1}\times\mathbf{S}_i),
\end{equation}
where $\mathbf{D}_{i,j}$ denotes the DM vector connecting sites $i$ and $j$, which obeys the antisymmetric relation $\mathbf{D}_{j,i}=-\mathbf{D}_{i,j}$, and $\mathbf{S}_i$ represents the spin vector at site $i$.  By applying vector identities, the effective field at the atomic site $i$ is given by
\begin{equation}
    \mathbf{h}_i^{\text{DM}}=\mathbf{D}_{i-1,i}\times\mathbf{S}_{i-1}-\mathbf{D}_{i,i+1}\times\mathbf{S}_{i+1}.\label{eq:field}
\end{equation}
In the continuum approximation, the spin and DM vectors can be expanded in cylindrical coordinates
\begin{eqnarray}
    \mathbf{S}_{i\pm 1}\simeq \mathbf{S}_i\pm\frac{\partial\mathbf{S}_i}{\partial z}\delta z\pm\frac{\partial\mathbf{S}_i}{\partial \varphi}\delta\varphi,\label{eq:spin}\\
    \mathbf{D}_{i,i+1}\simeq\mathbf{D}_{i-1,i}+\frac{\partial\mathbf{D}_{i-1,i}}{\partial z}\delta z\label{eq:dmi}.
\end{eqnarray}
Here, $\mathbf{D}_{i,j}$ varies along the $z$-axis. Then, substituting Eqs.~$\eqref{eq:spin}$ and $\eqref{eq:dmi}$ into Eq.~$\eqref{eq:field}$ and keeping only first-order variations, the DM field at site $i$ is given by

\begin{widetext}

\begin{equation}
    \mathbf{h}_i^{\text{DM}}=-2\mathbf{D}_{i-1,i}\times\frac{\partial\mathbf{S}_i}{\partial z}\delta z-2\mathbf{D}_{i-1,i}\times\frac{\partial\mathbf{S}_i}{\partial \varphi}\delta \varphi-\frac{\partial\mathbf{D}_{i-1,i}}{\partial z}\times\mathbf{S}_i\delta z.
\end{equation}
Taking the interfacial DMI model $\mathbf{D}_{i,j}\perp(\mathbf{r}_j-\mathbf{r}_i)$ \cite{Crepieux98}, the effective DM field in the micromagnetic approximation is given by
\begin{equation}
    \mathbf{H}^\text{DM}\simeq-\frac{2D(z)}{\mu_0M_\text{s}^2}\left[\left(\frac{\partial M_z}{\partial z}+\frac{M_\rho}{\rho}+\frac{1}{\rho}\frac{\partial M_\varphi}{\partial\varphi}\right)\boldsymbol{\hat{\rho}}+\frac{1}{\rho}\left(M_\varphi-\frac{\partial M_\rho}{\partial\varphi}\right)\boldsymbol{\hat{\varphi}}-\frac{\partial M_\rho}{\partial z}\mathbf{\hat{z}}\right]-\frac{1}{\mu_0M_\text{s}^2}\frac{\partial D(z)}{\partial z}\left[M_z\boldsymbol{\hat{\rho}}-M_\rho\mathbf{\hat{z}}\right].\label{eq:fieldfinal}
\end{equation}
%
After performing the appropriate vector manipulations, the effective field can be expressed as
\begin{equation}
    \mathbf{H}^{\text{DM}}=-\frac{2D(z)}{\mu_0M_\text{s}^2}\left[\boldsymbol{\hat{\rho}}(\nabla\cdot\mathbf{M})-\nabla(\boldsymbol{\hat{\rho}}\cdot\mathbf{M})+\frac{M_\varphi}{\rho}\boldsymbol{\hat{\varphi}}\right]-\frac{1}{\mu_0M_\text{s}^2}\frac{\partial D(z)}{\partial z}\left[\boldsymbol{\hat{\varphi}}\times\mathbf{M}\right],
\end{equation}
\end{widetext}
where the components of the magnetization vector, expressed in terms of the local orthonormal basis, are given by
\begin{eqnarray*}
M_\rho&=&\sum_{n}m_{\rho,G_n}e^{i[(k+G_n)z+\ell\varphi-\omega t]},\\
M_\varphi&=&M_\text{s}\,\mathcal{C}\sin\theta-\sum_{n}m_{\chi,G_n} e^{i[(k+G_n)z+\ell\varphi-\omega t]} \cos\theta,\\
M_z&=&M_\text{s}\cos\theta+ \sum_{n}m_{\chi,G_n} e^{i[(k+G_n)z+\ell\varphi-\omega t]}\mathcal{C}\sin\theta.
\end{eqnarray*}
Observe that, in Eq.~\eqref{eq:fieldfinal}, in the limit of large inner radii one obtains $1/\rho \to 0$ and $\boldsymbol{\hat{\rho}} \to \mathbf{\hat{y}}$, thereby recovering the expression corresponding to a chiral planar magnonic crystal \cite{Gallardo19c}
\begin{equation}
    \mathbf{H}^\text{DM}=\frac{2D(z)}{\mu_0M_\text{s}^2}\!\left(\!\frac{\partial m_y}{\partial z}\mathbf{\hat{z}}-\frac{\partial m_z}{\partial z}\mathbf{\hat{y}}\!\right)
    -\frac{\partial D(z)}{\partial z}
    \!\left(\!\frac{m_z\mathbf{\hat{y}}-m_y\mathbf{\hat{z}}}{\mu_0M_\text{s}^2}\!\right).
\end{equation}

\section{Matrix elements}
\label{AppBB}

To construct the dynamical matrix, the linearized effective fields are projected onto the local transverse basis ${\hat{\boldsymbol{\chi}},\hat{\boldsymbol{\rho}}}$ and the Bloch harmonics $\exp[i(k+G_n)z]$. The indices $n$ and $n'$ label the row and column reciprocal-space harmonics, respectively. With these conventions, and assuming that only the DMI interaction is periodic, the four blocks of the dynamical matrix in Eq.~\eqref{eq:eigen_tube} are
\begin{widetext}
    \begin{eqnarray}
\mathcal{T}_{\chi\chi}^{nn'}&=&\left[\frac{4\pi i\ell\lambda_\text{ex}^2\cos\theta}{S}\ln\frac{R}{R_{\rm i}}+\mathcal{N}_{\rho\chi}^{\text{dip}}\right]\delta^{n}_{n'}+\frac{2iD_{G_{n-n'}}}{\mu_0M_\text{s}^2}\left[\frac{2\ell\cos\theta}{R+R_{\rm i}} -\left(\frac{G_{n+n'}}{2}+k\right)\mathcal{C}\sin\theta\right]\label{Txx},\\
\mathcal{T}_{\chi\rho}^{nn'}&=&
\left\{\frac{H_0}{M_\text{s}}+\frac{H_{\rm u}}{M_{\rm s}}\cos^2\theta+\left[(G_{n'}+k)^2+\frac{2\pi}{S}(\ell^2+\cos^2\theta)\ln\frac{R}{R_{\rm i}}\right]\lambda_\text{ex}^2+\mathcal{N}_{\rho\rho}^{\rm dip}\right\}\delta^{n}_{n'}+\frac{4D_{G_{n-n'}}\cos^2\theta}{\mu_0M_\text{s}^2\left(R+R_{\rm i}\right)},\\
\mathcal{T}_{\rho\chi}^{nn'}&=&-\left\{\frac{H_0}{M_\text{s}}+\frac{H_{\rm u}}{M_{\rm s}}\cos 2\theta+\left[(G_{n'}+k)^2+\frac{2\pi}{S}(\ell^2+\cos2\theta)\ln\frac{R}{R_{\rm i}}\right]\lambda_\text{ex}^2+\mathcal{N}_{\chi\chi}^{\rm dip}\right\}\delta^{n}_{n'}-\frac{4D_{G_{n-n'}}\cos2\theta}{\mu_0M_\text{s}^2\left(R+R_{\rm i}\right)}
    \end{eqnarray}
and 
\begin{eqnarray}
    \mathcal{T}_{\rho\rho}^{nn'}&=&-\left[\frac{4\pi i\ell\lambda_\text{ex}^2\cos\theta}{S}\ln\frac{R}{R_{\rm i}}+\mathcal{N}_{\chi\rho}^{\text{dip}}\right]\delta^{n}_{n'}+\frac{2iD_{G_{n-n'}}}{\mu_0M_\text{s}^2}\left[\frac{2\ell\cos\theta}{R+R_{\rm i}}-\left(\frac{G_{n+n'}}{2}+k\right)\mathcal{C}\sin\theta\right]\label{Trr}.
\end{eqnarray}
\end{widetext}
The structure of Eqs.~\eqref{Txx}--\eqref{Trr} separates the contributions that preserve a reciprocal-space harmonic from those that couple different harmonics. Terms proportional to the Kronecker delta $\delta_{nn'}$ are diagonal in reciprocal space and arise from the spatially uniform applied field, anisotropy, exchange, and dipolar contributions. By contrast, the terms proportional to $D_{G_n-G_{n'}}$ originate from the periodically modulated DMI and couple Bloch components whose reciprocal vectors differ by $G_n-G_{n'}$. This coupling provides the Bragg scattering that gives rise to the magnonic bands. For uniform DMI, $D_{G_n-G_{n'}}=D\delta_{nn'}$, and the different reciprocal-space harmonics decouple.

The diagonal blocks $\mathcal{T}_{\chi\chi}^{nn'}$ and $\mathcal{T}_{\rho\rho}^{nn'}$ contain the chiral contributions associated with curvature, dipolar coupling, and DMI, whereas $\mathcal{T}_{\chi\rho}^{nn'}$ and $\mathcal{T}_{\rho\chi}^{nn'}$ contain the nonchiral contributions, including the curvature-induced exchange corrections and anisotropy-like DMI terms~\cite{Sheka22,Sheka22b}.
The quantities $\mathcal{N}_{\rho\rho}^{\rm dip}$, $\mathcal{N}_{\rho\chi}^{\rm dip}$, $\mathcal{N}_{\chi\rho}^{\rm dip}$, and $\mathcal{N}_{\chi\chi}^{\rm dip}$ are dimensionless dipolar matrix elements obtained by averaging the magnetostatic Green function over the nanotube cross section. For the $n$th Bloch harmonic, we define $q_n=k+G_n$ and use the cylindrical Green-function kernel
$$
\mathcal{G}_{\ell,n}(\rho,\rho')
=
2\pi I_\ell\!\left(|q_n|\rho_<\right)
K_\ell\!\left(|q_n|\rho_>\right),
$$
where $\rho_<=\min(\rho,\rho')$, $\rho_>=\max(\rho,\rho')$, and $I_\ell$ and $K_\ell$ are modified Bessel functions of order $\ell$. For compactness, the dependence on $n$ is suppressed below, so that $\mathcal{G}_{\ell,n}(\rho,\rho')$ is denoted simply by $\mathcal{G}_{\ell}(\rho,\rho')$. The dipolar matrix elements are then given by
\begin{widetext}
    \begin{eqnarray}
    \mathcal{N}^{\rm dip}_{\rho\rho}&=&\frac{1}{S}\int_{R_{\rm i}}^{R}\partial_\rho\left[\mathcal{G}_{\ell}(\rho,\rho')\rho'\vert ^{\rho'=R}_{\rho'=R_{\rm i}}-\int_{R_{\rm i}}^{R} \mathcal{G}_{\ell}(\rho,\rho') d\rho'\right]\rho d\rho\label{Nrhorho},\\
    \mathcal{N}^{\rm dip}_{\rho\chi}&=&\frac{i}{S}\int_{R_{\rm i}}^{R}\left((G_n+k)\,\mathcal{C}\sin\theta-\frac{\ell\cos\theta}{\rho}\right)\left[\mathcal{G}_{\ell}(\rho,\rho')\rho'\vert ^{\rho'=R}_{\rho'=R_{\rm i}}-\int_{R_{\rm i}}^{R}\mathcal{G}_{\ell}(\rho,\rho')d\rho'\right]\rho d\rho,\\
    \mathcal{N}^{\rm dip}_{\chi\rho}&=&-\frac{i}{S}\int_{R_{\rm i}}^{R}\partial_\rho\left[\int_{R_{\rm i}}^{R}\left((G_n+k)\,\mathcal{C}\sin\theta-\frac{\ell\cos\theta}{\rho'}\right) \mathcal{G}_{\ell}(\rho,\rho')\rho'd\rho'\right]\rho d\rho,
\end{eqnarray}
and 
\begin{equation}
    \mathcal{N}_{\chi\chi}^{\rm dip}=\frac{1}{S}\int_{R_{\rm i}}^{R}\left((G_n+k)\,\mathcal{C}\sin\theta-\frac{\ell\cos\theta}{\rho}\right)\left[\int_{R_{\rm i}}^{R}\left((G_n+k)\,\mathcal{C}\sin\theta-\frac{\ell\cos\theta}{\rho'}\right)\mathcal{G}_{\ell}(\rho,\rho')\rho'd\rho'\right]\rho d\rho.\label{Nchichi}
\end{equation}
\end{widetext}
The radial integrations in Eqs.~\eqref{Nrhorho}--\eqref{Nchichi} account for the finite thickness of the nanotube and are normalized by its annular
cross-sectional area $S$.

%
%

%


\end{document}